\documentclass[twocolumn,prb,amsmath,amssymb,showpacs,superscriptaddress]{revtex4-1}

\usepackage{graphicx}
\usepackage{epstopdf}
\usepackage{hyperref}
\usepackage{url}
\pdfoutput=1
\begin{document}
\newcommand{\ben}{\begin{equation}}
\newcommand{\een}{\end{equation}}

\title{Electric-field induced domain-wall dynamics: depinning and chirality switching}

\author{Pramey Upadhyaya}  
\affiliation{Department of Electrical Engineering, University of California, Los Angeles, California 90095, USA}

\author{Ritika Dusad}
\affiliation{Department of Physics and Astronomy, University of California, Los Angeles, California 90095, USA}

\author{Silas Hoffman}
\affiliation{Department of Physics and Astronomy, University of California, Los Angeles, California 90095, USA}

\author{Yaroslav Tserkovnyak}
\affiliation{Department of Physics and Astronomy, University of California, Los Angeles, California 90095, USA}

\author{Juan G. Alzate}
\affiliation{Department of Electrical Engineering, University of California, Los Angeles, California 90095, USA}

\author{Pedram Khalili Amiri}
\affiliation{Department of Electrical Engineering, University of California, Los Angeles, California 90095, USA}

\author{Kang L. Wang}
\affiliation{Department of Electrical Engineering, University of California, Los Angeles, California 90095, USA}

\begin{abstract}

We theoretically study the equilibrium and dynamic properties of nanoscale magnetic tunnel junctions (MTJs) and magnetic wires, in which an electric field controls the magnetic anisotropy through spin-orbit coupling. By performing micromagnetic simulations, we construct a rich phase diagram and find that, in particular, the equilibrium magnetic textures can be tuned between N\'{e}el and Bloch domain walls in an elliptical MTJ. Furthermore, we develop a phenomenological model of a quasi-one-dimensional domain wall confined by a parabolic potential and show that, near the  N\'{e}el-to-Bloch-wall transition, a pulsed electric field induces precessional domain-wall motion which can be used to reverse the chirality of a N\'{e}el wall and even depin it. This domain-wall motion controlled by electric fields, in lieu of applied current, may provide a model for ultra-low-power domain-wall memory and logic devices.

\end{abstract}


\maketitle

\section{Introduction}
Spin-transfer torque (STT) \cite{Slonczewski1996L1,*Berger96} has emerged as an effective mechanism to control memory and logic devices such as magnetic tunnel junction (MTJ)-based nonvolatile magnetic random-access memory \cite{Chen2010} and domain-wall-based race-track memory.\cite{Parkin11042008} However, the spin-polarized currents required for these devices are notoriously high and therefore energy inefficient, prompting the search for alternate low-power electric control. One promising alternative is voltage-controlled magnetic anisotropy (VCMA), in which an applied voltage induces an excess charge at the interface between the tunnel-barrier and transition-metal ferromagnet in an MTJ. This modifies the occupation and, possibly, the character of ``$d$-like'' bands in the transition metal, which, via spin-orbit coupling, modifies the anisotropy perpendicular to the insulator-ferromagnetic interface.\cite{maruyamaNATNANO09,*Duan2008,*Nakamura2008} Controlling magnetic anisotropy by applying an electric field alleviates the use of power-hungry currents in order to manipulate MTJs. For instance, VCMA has been shown experimentally and theoretically to assist switching by externally applied magnetic field,\cite{shiota2009,*wangNATMAT11,*alzate12} induce precessional switching,\cite{shiotaNATMAT11} induce ferromagnetic resonance,\cite{zhuPRL12} and provide control of the threshold switching current.\cite{2012arXiv1209.0962L} On the other hand, while VCMA has been shown to modulate domain-wall velocity \cite{Shellekens2012,*Bauer}and nucleation,\cite{bernand}to the best of our knowledge,  inducing domain-wall motion purely by electric field has neither been theoretically proposed nor experimentally observed.

\begin{figure}[h]
\centering
\includegraphics[width=8cm,height=5cm]{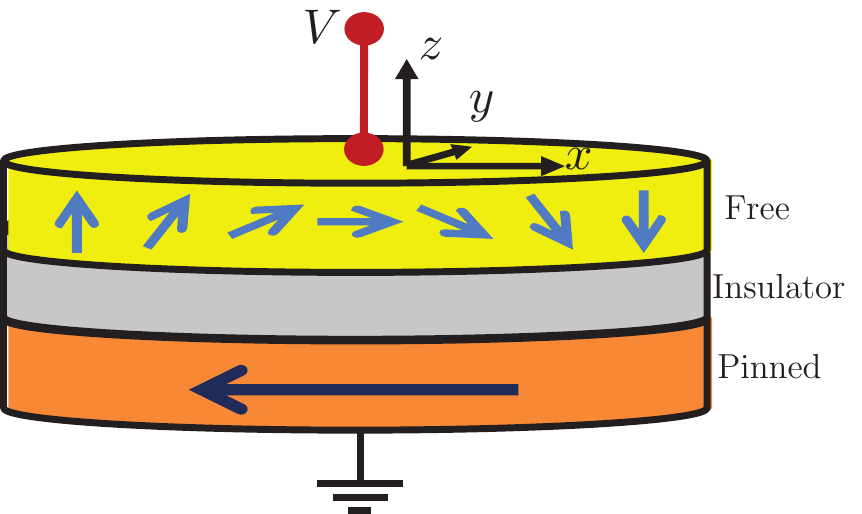}
\caption{The MTJ stack modeled by micromagnetic simulations: the free layer/insulator interface gives rise to an interfacial anisotropy $K^\perp$ with vertical easy axis, controlled by a voltage, $V$.\cite{maruyamaNATNANO09,*Duan2008,*Nakamura2008}Pinned layer is kept fixed to point along -$x$.}
\label{geom}
\end{figure}

In this paper, by determining the static micromagnetic configurations and their dynamic response to electric fields, we provide a method for inducing domain-wall motion purely by electric fields. To analyze the equilibrium magnetic texture, we consider a ferromagnet(F)$\mid$insulator (I)$\mid$F heterostructure with an elliptical cross section, i.e., an MTJ, where one of the ferromagnetic layers is pinned and the other free. See Fig.~\ref{geom}. The presence of inhomogeneous stray fields from the pinned layer are essential to decrease the energy cost for out-of-plane magnetization and therefore encourage out-of-plane texture. By performing micromagnetic simulations, we find that a domain wall can be stabilized when the perpendicular anisotropy, controlled by applied voltage, overcomes the out-of-plane demagnetization field. Informed by micromagnetic simulations, we study the VCMA-induced dynamics of this confined domain-wall by analytically modeling it as a quasi-one-dimensional magnetic wire. We find that VCMA induces precessional dynamics which can be exploited to switch the chirality of a N\'{e}el domain wall or force it to escape confinement.

\section{Equilibrium configurations}

We consider a device geometry, shown in Fig. 1, which consists of an MTJ with a free ferromagnetic layer and a pinned ferromagnetic layer, pointing along $-x$,  separated by an insulator having thicknesses $t_F$, $t_P$, and $t_I$, respectively. The cross section of the heterostructure is an ellipse where the semimajor axis is along the $x$ direction and the semiminor axis is along the $y$ direction.

\begin{figure}[h]
\includegraphics[width=8cm,height=5.4cm]{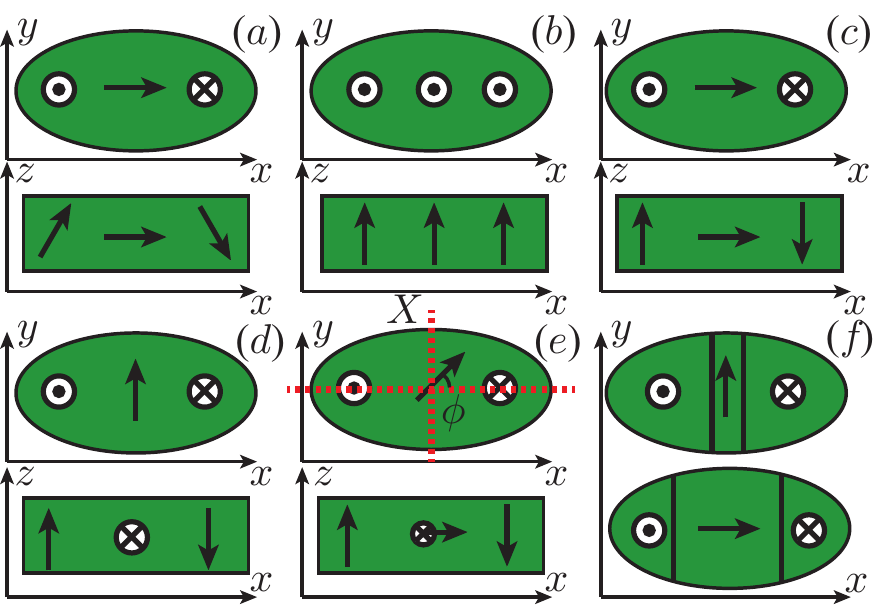}
\centering
\caption{Schematics of the relevant equilibrium configurations: (a) in-plane state with canted edges, (b) perpendicular state, (c) N\'{e}el domain wall, (d) Bloch domain wall, and (e) the intermediate domain wall where $\phi$ is the angle of magnetization at the center of the ellipse with respect to the $x$ axis. The lower right hand frame (f) depicts the domain wall width, i.e. the central region delineated with vertical lines, much larger (top) and smaller (bottom) than the semiminor axis of the ellipse corresponding to a N\'{e}el and Bloch wall, respectively.}
\label{state_schem}
\end{figure}

There are several magnetic configurations necessary to characterize the equilibrium states of our system as a function of perpendicular anisotropy. We schematically draw them in Fig.~\ref{state_schem}. The magnetization can be primarily in the plane of the ellipse along the $x$ axis [see Fig.~\ref{state_schem}(a)], however the texture near the edges of the ellipse tend to follow the stray fields of the pinned layer and cant out of plane at angle $\theta_c$. If the perpendicular anisotropy is large enough, the magnetization points entirely along the $z$ axis [see Fig.~\ref{state_schem}(b)]. Fig.~\ref{state_schem}(c) and (d) depict N\'{e}el and Bloch domain walls, respectively, where the magnetic domains are along the $z$ axis. A third domain wall, we call Bloch-like, in which the magnetization in the center of the domain wall is in-plane and aligned at an angle $\phi$ away from the $x$ axis [see Fig.~\ref{state_schem}(e)]. A N\'{e}el (Bloch) domain wall is a limiting example of a Bloch-like wall with $\phi=0^\circ$ ($\phi=180^\circ$) and can smoothly interpolate between N\'{e}el and Bloch domain walls, as  follows: When the domain-wall width ($\delta_W$), schematically drawn in Fig.~\ref{state_schem}(f) and controlled by perpendicular anisotropy, is large the energy is minimized due to dipolar interactions when the central region points along the $x$ axis. As $\delta_W$ is decreased, becoming narrower than the semiminor axis of the ellipse, the dipole energy in the N\'{e}el wall increases, forcing the magnetization at the center of the wall to orient away from the $x$ axis to a finite value of $\phi$.\cite{jung,*koyama} 

\begin{figure}[h]
\centering
\includegraphics[width=8cm,height=6.5cm]{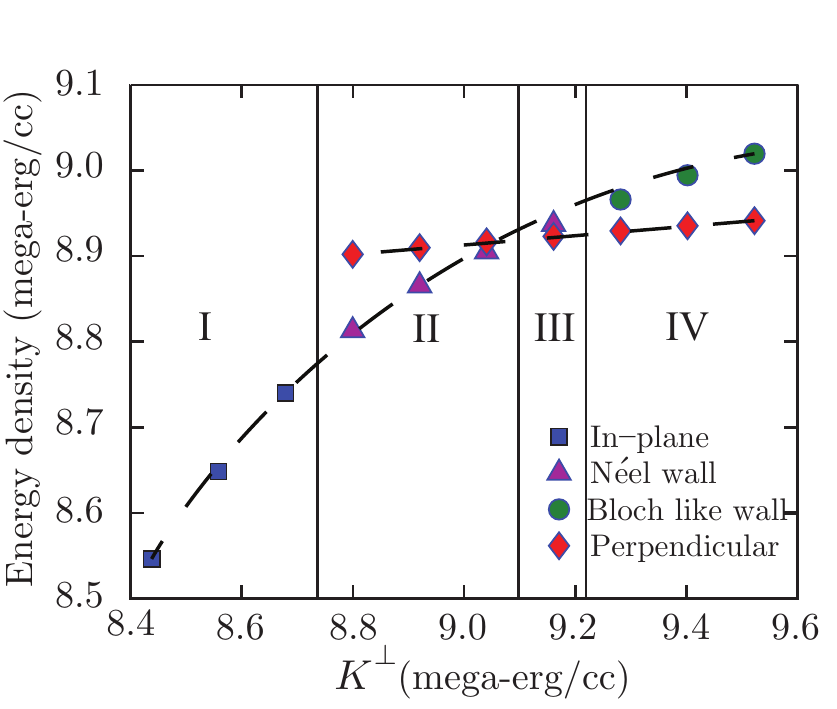}
\caption{The dependence of energy of equilibrium configurations on the interfacial perpendicular anisotropy: the vertical lines mark approximate boundaries between regions marked I-IV. Region I has only in-plane state with canted edges (schematically shown in Fig.~\ref{state_schem}a) as a stable state, region II  has both N\'{e}el  domain wall (schematically shown in Fig.~\ref{state_schem}c) and pependicular (monodomain) states, with domain walls having lower energy. In region III perpendicular states become lower in energy whereas in region IV the N\'{e}el domain wall starts transforming into Bloch form (referred as Bloch-like wall and schematically depicted in Fig.~\ref{state_schem}e). }
\label{eq_energy}
\end{figure}

To develop a precise numerical model using micromagnetic simulations we take the magnetization saturation $M_s=1200$~emu/cm$^3$ and exchange stiffness $A=2$~$\mu$erg/cm. The perpendicular anisotropy is assumed to be of the form, $K^\perp=k^\perp/t_F$, with $k^\perp=1.35$~erg/cm$^2$. These choices of material parameters are typical for a CoFeB/MgO interface. \cite{ikedanm10} For equilibrium simulations we control the perpendicular anisotropy by varying the free layer thickness. The major and minor axes are $300$ nm and $100$ nm respectively, while the thickness $t_I=2$ nm and $t_P=2.5$ nm. All the simulation results presented in this paper are obtained for zero temperature and are performed using the LLG Micromagnetic Simulator.\cite{LLG} The phase diagram was constructed by initializing the free magnetic layer along $x$ or along the $z$ direction and allowing the system to relax to equilibrium (possibly resulting in two different configurations) for values of $t_F$ ranging between $1.4$~nm and $1.6$~nm. Although increasing the thickness will change the demagnetization energy and influence of the stray fields, we do not expect it to significantly change our results. Energy of these equilibrium configurations are plotted as a function of perpendicular anisotropy in Fig.~\ref{eq_energy}.

When the thickness is large $t_F\gtrsim1.5$~nm corresponding to $K^\perp\lesssim8.7$~Merg/cm$^3$, the magnetization is primarily in-plane with canted edges and we identify the texture with the schematic in Fig.~\ref{state_schem}(a). As the perpendicular anisotropy increases, $\theta_c$ increases until $K^\perp\approx8.7$~Merg/cm$^3$ when the out-of-plane perpendicular anisotropy field overcomes the demagnetization field resulting in two stable magnetic configurations. The ground state is a N\'{e}el domain wall while we find an additional metastable  state with  magnetization perpendicular to the interface. The domain wall width decreases monotonically with increasing perpendicular anisotropy and at $K^\perp\approx9.1$~Merg/cm$^3$ the energy of the N\'{e}el domain-wall configuration surpasses that of the perpendicular state, which is now the most energetically favorable configuration. Although the perpendicular state remains the ground state, when $K^\perp\approx9.2$~Merg/cm$^3$ the metastable N\'{e}el state undergoes a second-order phase transition to the Bloch-like wall with symmetry broken between $+\phi$ and $-\phi$. Further increasing the perpendicular anisotropy shrinks the domain wall, thus increasing $\phi$. This control of $\phi$ with perpendicular anisotropy will be central in the discussion of dynamics in the following sections.
 
\section{Electric-field induced domain-wall dynamics }

In this section we consider the dynamic response of N\'{e}el wall configuration to voltage. First we focus on the micromagnetics governed by the Landau-Lifshitz-Gilbert (LLG) equation\cite{landauBK80,*gilbertIEEEM04} then, informed by these numerical results, we develop a phenomenological quasi-one-dimensional model, based on the Walker ansantz,\cite{walker1974} and find qualitative agreement between the two.

\begin{figure}[h]
\includegraphics[width=8cm,height=6.23cm]{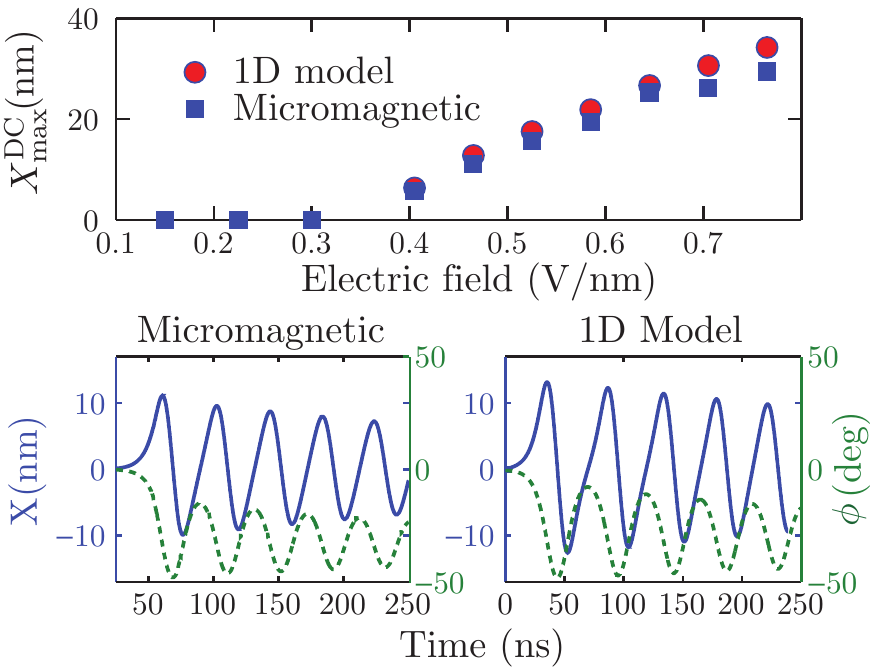}
\centering
\caption{Comparison between micromagnetics and 1D Model: (top panel) the maximum deviation of domain wall from the center, $X_{\rm max}^{\rm DC}$, obtained for electric field applied in a step-like fashion, (bottom panel) the dynamical approach to a Bloch-like wall from the initial configuration of a N\'{e}el wall, for same electric field magnitude in micromagnetic simulations (bottom left) and 1D Model (bottom right). Here solid lines depict the position ($X$) while the dashed lines show the angle ($\phi$) of the domain wall.}
\label{micro_vs_model}
\end{figure}

\textit{Micromagnetics}|The dynamic simulations are performed using the LLG equation: 
\ben
\dot{\textbf{m}}=-\gamma \textbf{m}\times \textbf{H}^{\rm{eff}}+ \alpha \textbf{m}\times\dot{\textbf{m}}\,,
\label {LLG}
 \een
where $\textbf{m}$ is the unit vector pointing along the position-and time-dependent magnetization configuration of the free layer, $\alpha$ is the Gilbert damping and $\gamma$ is the gyromagnetic ratio. The effective field, $\textbf{H}^{\rm{eff}}$, includes contributions from demagnetization field, exchange field and anisotropy field. In particular, VCMA torque enters due to electric field dependent anisotropy field which is taken into account by making $K^\perp=(k^\perp+\zeta E)/t_F$, in line with recent experiments, where $E$ is the electric field. The magneto-electric coupling coefficient is taken to be $\zeta=10^{-8}$~erg/V$\cdot$cm (close to the experimental\cite{endo2010,wangNATMAT11,zhuPRL12} and theoretical value\cite{niranjan}), $\alpha=0.01$ and $t_F=1.5$~nm, while leaving all other parameters as in the previous section. Because we are primarily interested in low dissipation devices, we include the Gilbert damping but do not include other dissipative torques (e.g. spin transfer torque) allowed by symmetry.\cite{Silas}  We initialize the system so that the free layer is a N\'{e}el domain wall which, for these parameters, is the ground state of our system. To induce motion we turn on an electric field abruptly to a non zero value and allow the domain wall to equilibrate to a new equilibrium state. We summarize the resultant motion in Fig.~\ref{micro_vs_model}. 

The top panel of Fig.~\ref{micro_vs_model} shows the maximum deviation of the domain wall, $X^{\rm{DC}}_{\rm{max}}$, from its equilibrium at the center of the ellipse as a function of the electric field magnitude. Below some critical value of electric field $E_c \approx 0.35$~V/nm, domain-wall motion is not induced, i.e.  $X^{\rm{DC}}_{\rm{max}}=0$. However when the magnitude of the electric field exceeds $E_c$, the domain wall position and angle oscillate, differing by a $\pi/2$ phase, as shown in the lower left hand panel of Fig.~\ref{micro_vs_model}. Note that the time axis in the lower left hand panel begins at $\approx25$~ns because below this time the dynamics are dominated by high frequency modes wherein the long-time dynamics of the domain-wall position, $X$, and tilt, $\phi$ are ill-defined. Importantly, $E_c$ corresponds to a perpendicular anisotropy of $K^\perp\approx9.3$~Merg/cm$^3$ which is near the value of perpendicular anisotropy where the domain wall transitions from N\'{e}el to Bloch-like (see Fig.~\ref{eq_energy}). That is, below the critical field an electric field pulse will change the width of the N\'{e}el wall but not the type of domain wall. Above the critical field the domain wall undergoes a N\'{e}el-to-Bloch-like transition, which in turn results in domain-wall motion.
 
\textit{1D Model}| The micromagnetic simulations imply that the long-time domain-wall dynamics are dominated by the \textit{soft} modes which can be characterized by the generalized coordinates $X$ and $\phi$ [see Fig.~\ref{state_schem}(e)] . To construct a phenomenological model of the domain-wall motion, we begin by writing down the free energy density consistent with the symmetries of our device geometry which is invariant under a mirror transformation in the $yz$ plane as well as a mirror transformation in the $xz$ plane followed by a $\pi$ rotation around the $z$ axis. Because $X\rightarrow-X$ under the former and $\sin\phi\rightarrow-\sin\phi$ under the latter, only even contributions of $X$ and $\sin\phi$ are allowed:
\ben
\mathcal{F}(X,\phi)=\frac{C}{2} X^2+K_1\sin^2\phi+K_2\sin^4\phi\,.
\label{free}
\een
The phenomenological constant $C>0$ parameterizes the strength of a parabolic confining potential due to both inhomogeneous stray fields from the pinned layer and the finite size of the free layer. The relative strength of $K_1$ and $K_2$ determine the easy plane in which the domain wall rotates. Because a positive $K_2$ is required for a smooth transition of equilibrium states from N\'{e}el to Bloch wall, we henceforth take $K_2>0$ to match micromagnetic simulations. Higher order interactions and interactions of the form $X^4$ and $X^2\sin^2\phi$ are in principle allowed but not necessary to reproduce N\'{e}el and Bloch-like equilibrium configurations and are thus disregarded in the spirit of constructing a minimal model. To find the equation of motion for the generalized coordinates we use the one-dimensional Walker ansatz\cite{walker1974} which states, for ${\bf m}=\{\sin\theta\cos\varphi, \sin\theta\sin\varphi,\cos\theta\}$ with $\theta$ and $\varphi$ being polar and azimuthal angles, respectively, $\varphi=\phi$ and $\ln[\tan$($\theta/2)]=(x-X)/\delta_W$. The equations of motion for our generalized coordinates are immediately determined following Ref. \onlinecite{Bazaliy2008,*Thiele}:
\begin{align}
\alpha \frac {\dot X}{\delta_W}+\dot \phi&=-\gamma H_C \frac {X}{\delta_W}\,, \nonumber \\
\frac{\dot X} {\delta_W}-\alpha \dot \phi&=\left(\gamma H_1+\gamma H_2\sin^2\phi\right)\sin2\phi\,,
\label{eom}
\end{align}
where we have defined $H_C \equiv \delta_W C/M_S$, $H_1 \equiv K_1/M_S$ and $H_2 \equiv 2K_2/M_S$. Because $H_C>0$ and $H_2>0$, the stable fixed points of Eq.~(\ref{eom}), $(X_0,\phi_0)$, depend on the value of $H_1$. For all stable solutions, the domain wall is at the center of the ellipse ($X_0=0$) while the domain-wall angle takes the values
\ben
\sin\phi_0 = \left\{ \begin{array}{cc}
  0 \,, & ~~~H_1>0\\
  \pm\sqrt{\left|H_1\right|/H_2}\,, & ~~~0>H_1>-H_2\\
  \pm 1\,, & ~~~-H_2>H_1
       \end{array} \right.,
\label{phi_sol}
\een
which correspond to a N\'{e}el, Bloch-like, and Bloch domain wall, respectively. The symmetry of the structure admits two solutions for any value of $H_1$.

In this model the phenomenological anisotropy fields, $H_1$ and $H_2$, and confining field $H_C$, are implicitly dependent on the perpendicular anisotropy $K^{\perp}$ and hence functions of the electric field. We therefore identify $E_c$ with the value of electric field at which $H_1$ becomes negative. The observed domain-wall motion in micromagnetic simulations can then be explained as the following. If $E<E_c$, the equilibrium position is the same for both, when electric field is on and off, and thus the domain-wall tilt, as well as the position, is always in the equilibrium resulting in no motion. However, if $E>E_c$, $\phi_0$ for the case when the electric field is on is different from when the electric field is off. The approach of $\phi$ to the new equilibrium in turn induces domain-wall motion according to Eq.~(\ref{eom}).

In order to compare with micromagnetic simulations, we numerically integrate Eq.~(\ref{eom}), first extracting values for the phenomenological parameters: $H_C$ is found by comparing the displacement of the N\'{e}el domain-wall in the micromagnetics and the phenomenological model as a function of externally applied magnetic field along the $z$ axis. $H_1$ and $H_2$ can then be obtained by matching $\phi_0$ and the resonance frequency of the micromagnetics to the quasi-one-dimensional model. The dynamics of the phenomenological model are shown in the lower right hand panel of Fig.~\ref{micro_vs_model} showing the time dependence of a domain wall, initially in a N\'{e}el configuration, after an electric field is turned on. We find the phenomenological model qualitatively matches the micromagnetic simulations. The presented minimal model is also able to provide a reasonable quantitative estimate of $X^{\rm{DC}}_{\rm max}$, which will be useful in the following section. Full quantitative agreement with the micromagnetics would require the inclusion of additional terms in the free energy, e.g. $\sim$ $X^4$, $X^2\sin^2\phi$, which is outside the scope of the present paper.

\section{Depinning and Chirality Switching}
Next we study the dependence of the electric field pulse duration, $T$, on domain-wall motion. In contrast to the previous section, $T$ is much shorter than the time for the domain-wall to reach equilibrium. By exploiting the VCMA-induced dynamics we find that a short pulsed electric field can switch the domain-wall chirality, i.e. $\phi_0\rightarrow\phi_0+\pi$, or induce a maximum displacement greater than the quasiequilibrium response, i.e. $X>X^{\rm{DC}}_{\rm{max}}$, for the same pulse magnitude. We comment on the application of the former to magnetic memory storage and the latter to racetrack memory. 

\begin{figure}[h]
\centering
\includegraphics[width=8cm,height=7cm]{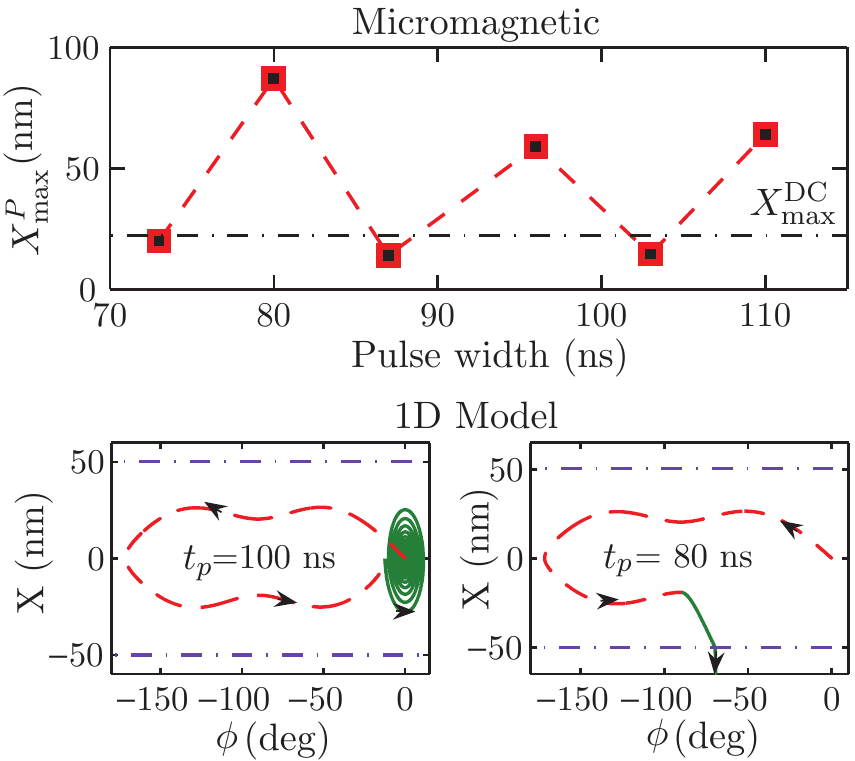}
\caption{Pulse width dependence of maximum domain wall deviation and depinning: (top panel)  oscillatory enhancement of maximum deviation of domain wall from center, obtained from micromagnetics for electric field $E=0.65$ V/nm. The dash-dotted horizontal line depicts the long-time maximum deviation ($X_{\rm max}^{\rm DC}$) for the same electric field as reference, (bottom panel) phase plot depicting the dependence of depinning on the pulse width as obtained within the 1D model for $E=0.65$V/nm. The extent of the parabolic potential $X_C=50$ nm is marked by the horizontal dash-dot lines while the trajectory for which the pulse is on and off is marked by a dashed and a solid line, respectively. Depinning occurs when pulse is turned off for $\phi$ close to 90$^\circ$ (bottom right panel) whereas no depinning occurs when pulse is turned off for $\phi$ close to 0$^\circ$ (bottom left panel)}
\label{depin}
\end{figure}

We perform micromagnetic simulations using the geometry and material parameters given in the previous section. In Fig.~\ref{depin} (upper panel), we plot the maximum displacement of the domain wall, $X^P_{\rm{max}}$, and find an oscillatory dependence as a function of pulse length. This is similar to pulsed switching of a monodomain by an applied magnetic field:\cite{schumacherPRL03} when an electric field pulse of magnitude $E>E_c$ is applied to our structure, the equilibrium domain-wall tilt changes from zero to a finite value which forces $\phi$ to oscillate around the new equilibrium. Because $\phi$ and $X$ are conjugate variables, the maximum domain-wall displacement increases with increasing tilt angle. Thus, turning off electric field pulse when $\phi$ is large enhances $X^P_{\rm{max}}$, relative to $X^{\rm{DC}}_{\rm{max}}$. Control of $X^P_{\rm max}$ by the duration of current pulses has likewise been demonstrated for the case of spin-torque induced domain-wall motion.\cite{Thomas2006}

\begin{figure}[h]
\centering
\includegraphics[width=8cm,height=6.9cm]{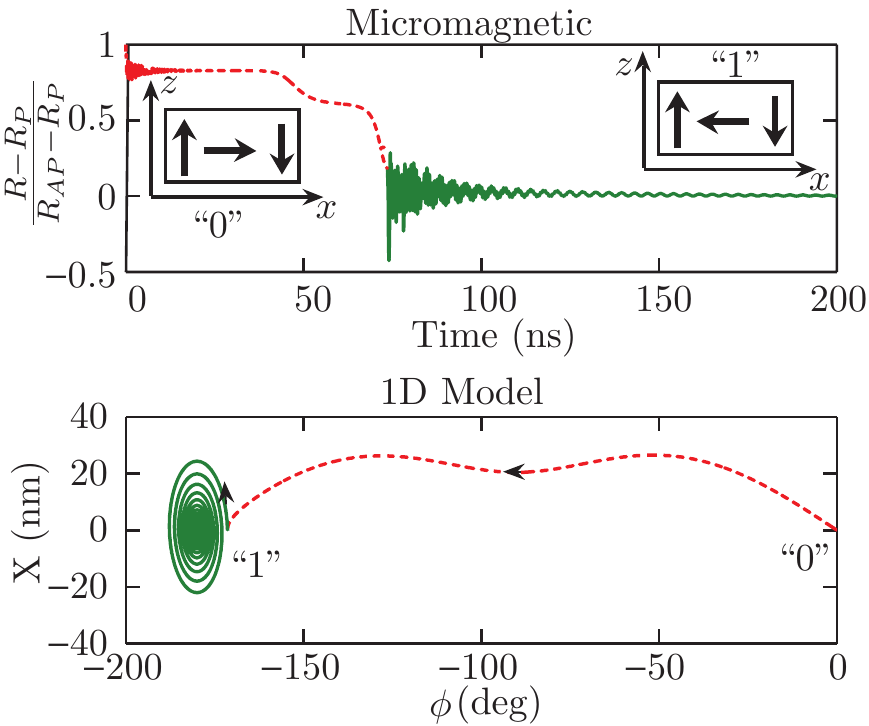}
\caption{Chirality switching: (top panel) switching from state ``0", having resistance of $R_{AP}$,  to state ``1", having a resistance of $R_P$ (along with the respective domain-wall configurations in free layer depicted schematically), obtained via micromagnetic simulation for electric field $E= 0.65$ V/nm, (bottom panel) switching trajectory in phase space from 1D model for the switching between state ``0"  and state ``1" for the same electric field.}
\label{chirality_switch}
\end{figure}

\textit{Chirality switching}| This control of the maximum amplitude of $\phi$ can be used to reverse the chirality of the domain wall. See Fig.~\ref{chirality_switch} (upper panel). In micromagnetic simulations we apply an electric field of magnitude 0.65~V/nm to a N\'{e}el wall initialized at $\phi_0=0$ which forces a new equilibrium value at $\phi_0 \approx 50^\circ$. As $\phi$ precesses around $\phi_0$, it goes past $90^\circ$ at which time the electric field is turned off and $\phi$ precesses towards the nearest N\'{e}el wall equilibrium, $\phi_0=180^\circ$, switching the chirality. Likewise, we reproduce chirality switching in the phenomonological model [Fig.~\ref{chirality_switch} (lower panel)]. Using the chirality to encode bit information, ``0'' when $\phi=0^\circ$ and ``1'' when $\phi=180^\circ$, tunnel magneto-resistance\cite{jullierePLA75} with state ``0" (having resistance $R_{AP}$) and state ``1" (having resistance $R_P$) to read the chirality, and VCMA to switch it, domain-wall memory switch can be constructed. Although current-induced spin torque has also been used to switch domain-wall chirality,\cite{PhysRevLett.101.107202} electric-field induced chirality switching provides a low-power attractive alternative. Moreover voltage control of chirality could provide an additional electrical knob to control recently observed efficient domain wall motion of chiral domain walls driven by spin Hall effect.\cite{emorinm13,*ryunnt13}

\textit{Depinning of domain walls }|We depart slightly from the free energy used in the previous sections, Eq.~\ref{free}, and consider, instead of a parabolic potential $CX^2/2$, a ``pinning parabolic potential'' in which 
\ben
\mathcal{F}[X] = \left\{ \begin{array}{cc}
  CX^2/2 \,, & ~~~\left|X\right|\leq X_c\\
  CX_c^2/2\,, & ~~~\left|X\right|>X_c
       \end{array} \right.,
\een
leaving all other parameters and the equation of motion, Eq.~\ref{eom}, unaltered. In this potential, if the displacement of the domain wall exceeds $X_c$ the domain wall is no longer localized to $X=0$, i.e. ``depins.'' Experimentally, this can be realized by depositing a pinned layer with length of the order of $ X_c$ on top of a nanowire. Taking $X_c=50$~nm and a pulse magnitude of $E=0.65$~V/nm, we use the modified quasi-one-dimensional model to study the effect of pulse width on depinning. When $T=100$~ns [Fig.~\ref{depin} (lower left panel)], the pulse turns off at $\phi\approx0$ and the domain wall returns to $X=0$. A pulse width of $T=80$~ns [Fig.~\ref{depin} (lower right panel)] ends at $\phi\approx90^\circ$, displacing the domain wall further than $X_c$, and depinning it. The ability to achieve depinning via pulsing electric field, in the absence of currents, could provide a major technological advantage in terms of power consumption in, for instance, racetrack memory.\cite{Parkin11042008} Another key difference from the corresponding spin-torque induced domain-wall motion is the direction of electric-field induced domain-wall motion. In the present case, the domain wall can move either to $+x$ or to $-x$ as the new equilibrium exists at both $\pm \phi_0$. A small $\pm y$ directed magnetic field, sufficient to overcome the thermal fluctuations, can break the aforementioned symmetry and favor $\pm \phi_0$, thus controlling the direction of motion.

\section{Summary}

By performing  micromagnetic simulations, we found  that in an MTJ with high perpendicular anisotropy a domain-wall state can be stabilized either as a ground or a metastable configuration (which is assisted by inhomogeneous stray fields emanating from the pinned layer), when the perpendicular anisotropy is strong enough to overcome the long-range out-of-plane dipolar fields. These domain walls transform smoothly from N\'{e}el  to Bloch structure with increasing value of perpendicular anisotropy. This transformation is driven by reduction of dipole energy stored inside the wall with decreasing domain-wall width, as the value of perpendicular anisotropy is increased. The N\'{e}el-to-Bloch-wall transormation also provides a route to induce domain-wall motion by electric field via VCMA. We developed a minimal phenomenological model explaining the oscillatory motion observed in micromagnetic simulations above a critical electric field, necessary to bring the N\'{e}el wall to the transition point of transforming into a Bloch wall. Finally, we use the predicted domain-wall motion to show that electric field pulses can be utilized to induce depinning of domain walls in magnetic wires (which in turn can be exploited in race-track like domain-wall memory devices), or achieve chirality switching (which can be used for making a domain-wall memory switch, storing information in the chirality of the domain wall or providing voltage control of chiral domain wall motion). The ability to manipulate domain walls via electric field could thus provide a path towards ultra low power magnetoelectric devices.

\begin{acknowledgements}
This work was supported in part by the DARPA program on nonvolatile logic and by the NSF Nanosystems Engineering Research Center for Translational Applications of Nanoscale Multiferroic Systems (TANMS), Grant No. 228481 from the Simons Foundation, and the NSF under Grant No. DMR-0840965.
\end{acknowledgements}

\end{document}